\documentclass[aps, prd, twocolumn, nofootinbib,]{revtex4-2}

\usepackage{amsmath}
\usepackage{amsfonts}
\usepackage{wasysym}
\usepackage{graphicx}
\usepackage{comment}
\usepackage{tensor}
\usepackage{cancel}
\usepackage[svgnames]{xcolor}

\usepackage[colorlinks = true, linkcolor = DarkBlue, citecolor = DarkBlue, urlcolor = DarkBlue, bookmarks = false]{hyperref}

\def\filetype{pdf}
\def\path{}

\allowdisplaybreaks[1]

\begin{document}


\title{Dynamical evolution and stability of quantum corrected\\ Schwarzschild black holes in semiclassical gravity}
\author{Ben Kain}
\affiliation{Department of Physics, College of the Holy Cross, Worcester, Massachusetts 01610, USA}

\begin{abstract}
\noindent
The Schwarzschild solution describes a classical static black hole in general relativity. When general relativity is extended by including semiclassical corrections in the form of a renormalized energy-momentum tensor, the horizon of the Schwarzschild black hole disappears and is replaced by a wormhole. We study the stability of this quantum corrected static Schwarzschild solution in semiclassical gravity by using it as the initial data of a dynamical evolution. We find that the quantum corrected solution is unstable and that the wormhole can expand or collapse when perturbed. In vacuum, the wormhole expands, but in the presence of even a small amount of classical matter, the wormhole collapses, forming a horizon and evolving to an evaporating black hole.
\end{abstract} 

\maketitle


\section{Introduction}

The Schwarzschild solution is the prototypical black hole solution in general relativity. It contains a horizon and a central singularity. When semiclassical corrections are included---which are arguably the simplest quantum corrections for extending general relativity---the horizon disappears and is replaced by a wormhole \cite{Fabbri:2005zn, Ho:2017joh, Ho:2017vgi, Arrechea:2019jgx}. A curvature singularity is located  beyond the wormhole throat, but it is no longer centrally located. Similar phenomena occur for the quantum corrected Reissner-Nordstr\"om \cite{Arrechea:2021ldl} and Einstein-Yang-Mills \cite{Kain:2025ygm} black holes.

Semiclassical corrections to general relativity are expected to remove the classical horizon for nonextremal \textit{static} black holes \cite{Berthiere:2017tms, Arrechea:2019jgx}. Static solutions are ubiquitous in the study of spherically symmetric gravity. However, it is important to remember that a static solution may be unstable with respect to time-dependent perturbations. If this is the case, then it is unlikely that the static solution will occur naturally and it certainly cannot occur as the final state in an evolving system. Whenever a static solution is found, an immediate and important question is whether or not it is stable.

A common method for determining stability is to perturb the static fields with time-dependent perturbations and then to linearize the system of equations with respect to the perturbations. This method can determine the linear stability of the static solution. Alternatively, the nonlinear stability of the static solution can be determined by dynamically evolving the system with the static solution supplying the initial conditions. If the static solution is unstable, the nonlinear evolution has the advantage of also determining the final state to which the static solution will evolve. In this work, we study the stability of the quantum corrected Schwarzschild black hole by dynamically evolving the static solution.

In semiclassical gravity \cite{Birrell:1982ix}, the field equations take the form
\begin{equation} \label{semiclassial field eqs}
G_{\mu\nu} = 8\pi \left( T_{\mu\nu} + \langle \widehat{T}_{\mu\nu} \rangle \right),
\end{equation}
where $G_{\mu\nu}$ is the Einstein tensor, $T_{\mu\nu}$ is the classical energy-momentum tensor, and $\langle \widehat{T}_{\mu\nu} \rangle$ is the semiclassical correction to general relativity, which is the expectation value of a renormalized energy-momentum tensor operator. We use the Polyakov approximation to compute $\langle \widehat{T}_{\mu\nu} \rangle$ \cite{Davies:1976ei, davies, Polyakov:1981rd}. This approximation was used to compute the quantum corrected static Schwarzschild solutions \cite{Fabbri:2005zn, Ho:2017joh, Ho:2017vgi, Arrechea:2019jgx} and has been used in dynamical evolutions \cite{Parentani:1994ij, Ayal:1997ab, Sorkin:2001hf, Hong:2008mw, Hwang:2010im}. Moreover, the Polyakov approximation allows for black hole evaporation \cite{Parentani:1994ij, Ayal:1997ab}.

From our simulations, we find that the quantum corrected Schwarzschild black hole is unstable and that the wormhole can exhibit both expansion and collapse. In the case of collapse, which occurs when the system is perturbed through the inclusion of classical matter, a horizon and a central singularity form and the system evolves to an evaporating black hole.

In dynamically evolving the quantum corrected Schwarzschild black hole, we are dynamically evolving a wormhole. As far as we are aware, very few wormholes have been studied dynamically. The Ellis-Bronnikov wormhole \cite{Ellis:1973yv, Bronnikov:1973fh} has been dynamically evolved by various groups \cite{Shinkai:2002gv, Gonzalez:2008xk, Doroshkevich:2008xm, Calhoun:2022xrw} and, recently, the Einstein-Dirac-Maxwell wormhole \cite{Blazquez-Salcedo:2020czn, Konoplya:2021hsm} has been dynamically evolved \cite{Kain:2023ore, Kain:2023ann}. We find similarities between the evolution of the quantum corrected Schwarzschild wormhole and the Ellis-Bronnikov wormhole. When relevant, we comment on these similarities.

In Sec.~\ref{sec:static}, we briefly review the static quantum corrected Schwarzschild black hole. In Sec.~\ref{sec:dynamical}, we describe our dynamical model and explain the numerical methods we use to solve it. In Sec.~\ref{sec:results}, we present our results and show that the static solution is unstable. We conclude in Sec.~\ref{sec:conclusion}. In the Appendix, we present tests of our code. Throughout we use units such that $c = G = \hbar = 1$.


\section{Static solutions}
\label{sec:static}

In this section, we review static quantum corrected Schwarzschild black holes in semiclassical gravity. Our main goal is to use these static solutions as initial data for the dynamical simulations we present in Sec.~\ref{sec:results}. Comprehensive studies of these static solutions are given in \cite{Fabbri:2005zn, Ho:2017joh, Ho:2017vgi, Arrechea:2019jgx}, to which we refer the reader for additional details. 

The quantum corrected solutions are found numerically. In computing solutions, different forms for the static spherically symmetric metric have been used, but a form we find convenient is
\begin{equation} \label{static metric}
ds^2 = e^{2\sigma(x)} (-dt^2 + dx^2) + r^2(x) d\Omega^2,
\end{equation}
where $x$ is the radial coordinate, $\sigma(x)$ and $r(x)$ parametrize the metric, $r(x)$ is the areal radius, and $d\Omega \equiv d\theta^2 + \sin^2\theta \, d\phi^2$.

The semiclassical field equations are given in (\ref{semiclassial field eqs}). In vacuum, the classical energy-momentum tensor vanishes, $T_{\mu\nu} = 0$. For the renormalized energy-momentum tensor, $\langle \widehat{T}_{\mu\nu} \rangle$, we use the Polyakov approximation and build it from the exact renormalized energy-momentum tensor computed in $1+1$ dimensions \cite{Davies:1976ei, davies, Polyakov:1981rd}. As is common, and for simplicity, we will present the renormalized energy-momentum tensor for $N$ massless scalar fields. In the Polyakov approximation, the renormalized energy-momentum tensor in spherically symmetric $3+1$ dimensions is taken to be
\begin{equation} \label{Polyakov approx}
\langle \widehat{T}_{ab} \rangle = \frac{3P}{r^2} \langle \widehat{T}_{ab} \rangle^{(\text{2D})},
\end{equation}
where $a,b$ are nonangular components, $\langle \widehat{T}_{\mu\nu} \rangle^{(\text{2D})}$ is the renormalized energy-momentum tensor in $1+1$ dimensions, and
\begin{equation}
P \equiv \frac{N}{12\pi} \ell_P^2, 
\end{equation}
where $\ell_P$ is the Planck length. In the units we are using, $\ell_P = 1$. The choice of the multiplicative factor $3P/r^2$ ensures that $\langle \widehat{T}_{\mu\nu} \rangle$ is conserved and that $\langle \widehat{T}_{\theta\theta} \rangle  = \langle \widehat{T}_{\phi\phi} \rangle = 0 $. 

For the metric in (\ref{static metric}), the renormalized energy-momentum tensor works out to
\begin{equation} \label{static RSET}
\begin{split}
\langle \widehat{T}_{tt} \rangle 
&= - \frac{P}{8 \pi r^2} (\sigma^{\prime \, 2} - 2\sigma'')
\\
\langle \widehat{T}_{xx} \rangle 
&= - \frac{P}{8 \pi r^2} \sigma^{\prime \, 2}
\end{split}
\end{equation}
and $\langle \widehat{T}_{\theta\theta} \rangle  = \langle \widehat{T}_{\phi\phi} \rangle = 0 $, where a prime denotes an $x$ derivative. From the semiclassical field equations in (\ref{semiclassial field eqs}), we can then derive the following second-order equations for the metric functions:
\begin{equation} \label{Schwarzschild counterpart static eqs}
\begin{split}
\sigma'' 
&= - \frac{1}{r}\left(2 r' \sigma'
+ \sigma^{\prime\,2} \frac{P }{r} \right)
\left(1 - \frac{P}{r^2} \right)^{-1}
\\
r'' 
&= \left(2 r' \sigma'
+ \sigma^{\prime\,2} \frac{P}{r} \right)
\left(1 - \frac{P}{r^2} \right)^{-1}.
\end{split}
\end{equation}
Note that these equations are divergent for $r\rightarrow \sqrt{P}$. This divergence follows from the choice of the multiplicative factor in (\ref{Polyakov approx}). This divergence will not play a role in this section because the static solutions we find will not include $r \leq \sqrt{P}$. This divergence will play a role in our dynamical simulations and we will comment on our interpretation of this divergence in the next section. For details on the derivation of (\ref{static RSET}) or (\ref{Schwarzschild counterpart static eqs}), see, for example, \cite{Barcelo:2011bb, Kain:2025ygm}.

One can show that as $x\rightarrow \infty$, the renormalized energy-momentum tensor drops off sufficiently quickly that the classical Schwarzschild spacetime, as parametrized by the ADM mass $M$, is a solution to (\ref{Schwarzschild counterpart static eqs}). This allows us to use the classical Schwarzschild spacetime for the outer boundary values when solving (\ref{Schwarzschild counterpart static eqs}). For classical Schwarzschild, with the metric in (\ref{static metric}), $x$ is the well-known tortoise coordinate. To solve for quantum corrected solutions, we proceed as follows. We choose a large value for $r$, which marks the outer boundary of the static solution, and a value for $M$. At the outer boundary, we have then
\begin{equation} \label{classiacl Schwarzschild}
\sigma = \frac{1}{2} \ln \left(1 - \frac{2M}{r} \right),
\qquad
x = r + 2M \ln \left( \frac{r}{2M} - 1 \right),
\end{equation}
and, upon taking $x$ derivatives,
\begin{equation}
\sigma' = \frac{M}{r^2},
\qquad
r' = \frac{r-2M}{r},
\end{equation}
which is the classical Schwarzschild solution. With these outer boundary values and given a value for $P$, we can integrate (\ref{Schwarzschild counterpart static eqs}) inward to obtain the quantum corrected Schwarzschild solution.

\begin{figure} 
\includegraphics[width=3in]{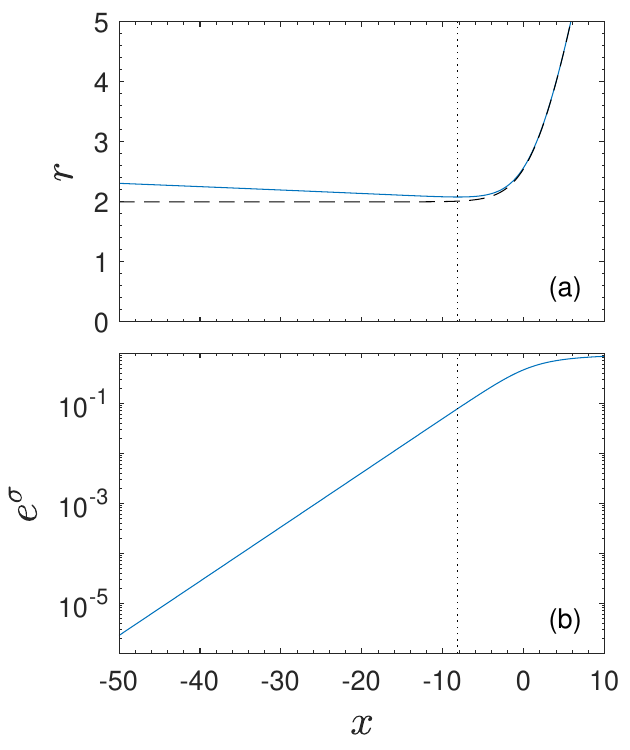} 
\caption{The solid lines of both plots display a quantum corrected Schwarzschild black hole with $M = 1$ and $P = 0.1$. (a) The areal radius, $r$, as a function of the radial coordinate, $x$. A minimum occurs at $x_\text{th} = -8.139$, indicating a wormhole throat with radius $r_\text{th} = 2.081$. The vertical dotted line indicates the position of the wormhole throat and the dashed line is the classical Schwarzschild solution, which is included for comparison. (b) The wormhole throat is horizonless since $e^{\sigma}$ is nonzero. It can be shown that the classical horizon has completely disappeared in the quantum corrected solution.}
\label{fig:static solutions}
\end{figure}

In Fig.~\ref{fig:static solutions}(a), we show the areal radius, $r$, as a function of $x$ for $M = 1$ and $P = 0.1$. Since we find a minimum for $r$, which is marked by the vertical dotted line, we have a wormhole structure. The value of $r$ at the minimum is the wormhole throat radius, $r_\text{th}$. For comparison, we show the classical solution, with $P = 0$, as the dashed line. In Fig.~\ref{fig:static solutions}(b), we show $e^{\sigma}$ as a function of $x$. Since $e^\sigma$ is nonzero, there is no horizon at the wormhole throat. It can be shown that as $x\rightarrow -\infty$, there exists a null curvature singularity at a finite proper distance \cite{Fabbri:2005zn, Ho:2017vgi, Arrechea:2019jgx}. As suggested by Fig.~\ref{fig:static solutions}(b), $e^\sigma$ continues to be nonzero and the classical horizon has completely disappeared in the quantum corrected solution. We refer to this as a wormhole, although we note that it may be more appropriate to refer to it as a ``wormhole structure" since the region $x\rightarrow - \infty$ is not asymptotically flat. Quantum corrected solutions with different values of $M$ and $P$ are qualitatively similar to Fig.~\ref{fig:static solutions}.

Increasing the mass $M$ increases the wormhole throat radius. Increasing $P$ does as well, but not by as much as when increasing $M$. For example, in Fig.~\ref{fig:static solutions}(a), $r_\text{th} = 2.081$, but for $M = 5$ and $P = 0.1$, the wormhole throat radius is $10.024$ and for $M = 5$ and $P = 1$ it is $10.186$. When we present dynamical evolutions of static solutions in Sec.~\ref{sec:results}, we cannot expect the relationship between the mass and wormhole throat radius that exists for the static solution to be maintained exactly during the evolution. However, we can assume an approximate relationship, with increasing (decreasing) mass corresponding to increasing (decreasing) radius. This will help us gain physical insight into how the radius increases or decreases.

It will be useful to define some language for referencing different regions of the quantum corrected spacetime. We'll refer to the region $x > x_\text{th}$ as ``outside" the wormhole. This is the region where we reside. The region $x < x_\text{th}$ is reached by passing through the wormhole throat. For convenience, we will refer to this region as ``inside" the wormhole, though more properly it is on the other side of the wormhole.

That semiclassical corrections cause the classical horizon of a Schwarzschild black hole to disappear and to be replaced with an asymmetric wormhole was first found in \cite{Fabbri:2005zn} and subsequently confirmed in \cite{Ho:2017joh, Ho:2017vgi, Arrechea:2019jgx}. Similar results have been found for the classical Reissner-Nordstr\"om \cite{Arrechea:2021ldl} and Einstein-Yang-Mills \cite{Kain:2025ygm} black holes. The question we ask is whether these static solutions, and hence wormholes like that shown in Fig.~\ref{fig:static solutions}, are stable with respect to time-dependent perturbations? If they are, then it may be possible for the wormhole to form naturally. On the other hand, if the static solutions are unstable, then it is impossible for the wormhole to form, at least as a final state. Further, if the static solutions are unstable, to what final state do they evolve?


\section{Dynamical solutions}
\label{sec:dynamical}

We turn now to our main goal, which is to dynamically evolve the quantum corrected Schwarzschild black hole. A dynamical evolution can determine if the static solutions discussed in Sec.~\ref{sec:static} are stable with respect to perturbations. If we find static solutions that are unstable, as will be the case, then a dynamical evolution can determine the final state to which the system will evolve. To perform the dynamical evolution we will develop a dynamical model and will use the static solutions found in Sec.~\ref{sec:static} as initial data.

For the dynamical model, we choose to work in double null coordinates and we parametrize the metric as
\begin{equation} \label{metric}
ds^2 = -e^{2\sigma(u,v)} du dv + r^2(u,v) d\Omega^2.
\end{equation}
The metric fields $\sigma$ and $r$ are the same metric fields in (\ref{static metric}), but of course now are no longer static, and the outgoing null coordinate $u$ and the ingoing null coordinate $v$ are defined with respect to the temporal and radial coordinates in (\ref{static metric}) in the usual way,
\begin{equation} \label{u v t x}
u = t - x,
\qquad
v = t + x.
\end{equation}
We choose to use double null coordinates for two important reasons. First, double null coordinates can straightforwardly incorporate the renormalized energy-momentum tensor in the Polyakov approximation \cite{Parentani:1994ij, Ayal:1997ab}.  Second, if a horizon forms, double null coordinates are well suited for computing the spacetime behind the horizon. We mention also that double null coordinates have been used to dynamically evolve wormholes \cite{Shinkai:2002gv, Doroshkevich:2008xm}. As mentioned in the Introduction, we will find similarities between the evolution of the quantum corrected Schwarzschild black hole and the evolution of the Ellis-Bronnikov wormhole  \cite{Shinkai:2002gv, Gonzalez:2008xk, Doroshkevich:2008xm, Calhoun:2022xrw}.

When numerically evolving a static solution, discretization error inherent in any numerical code acts as a small perturbation. Additionally, we will include at times an explicit perturbation in the form of a pulse of classical matter. For simplicity, we will use a real massless scalar field, $\phi$, for the pulse, described by the Lagrangian
\begin{equation} \label{Lagrangian}
\mathcal{L} = 
- \frac{1}{2} (\nabla_\mu \phi)(\nabla^\mu \phi),
\end{equation}
which we then minimally couple to gravity, $\mathcal{L} \rightarrow \sqrt{-\det(g_{\mu\nu})} \, \mathcal{L}$.


\subsection{Equations}
\label{sec:equations}

For the renormalized energy-momentum tensor, we continue to use the Polyakov approximation in (\ref{Polyakov approx}). For the metric in (\ref{metric}), we have \cite{Parentani:1994ij, Ayal:1997ab, Sorkin:2001hf, Hong:2008mw, Hwang:2010im}
\begin{equation} \label{RSET}
\begin{split}
\langle\widehat{T}_{uu}\rangle 
&= 
\frac{P}{4\pi r^2}
(\sigma_{,uu} - \sigma_{,u}^2)
\\
\langle\widehat{T}_{vv}\rangle 
&= 
\frac{P}{4\pi r^2}
(\sigma_{,vv} - \sigma_{,v}^2)
\\
\langle  \widehat{T}_{uv}\rangle
&= 
- \frac{P}{4\pi r^2}
\sigma_{,uv}
\end{split}
\end{equation}
and $\langle \widehat{T}_{\theta\theta} \rangle  = \langle \widehat{T}_{\phi\phi} \rangle = 0$, where we use the notation $\sigma_{,\mu} = \partial_\mu \sigma$ and similarly for other variables. Details on how these components are computed can be found in \cite{Fabbri:2005mw, Barcelo:2011bb}.

The only classical matter in the system is the scalar field described by the Lagrangian in (\ref{Lagrangian}). From this Lagrangian and the metric in (\ref{metric}), the equation of motion for the scalar field is the evolution equation
\begin{equation}  \label{phi evo}
\phi_{,uv}
= - \frac{1}{r} \left(r_{,u} \phi _{,v} +  r_{,v} \phi_{,u} \right)
\end{equation}
and the classical energy momentum tensor is
\begin{equation} 
\begin{split}
T_{uu} &= \phi_{,u}^2
\\
T_{vv} &= \phi_{,v}^2
\\
T_{uv} &= 0
\\
T_{\theta\theta} &= 2 r^2 e^{-2\sigma} \phi_{,u} \phi_{,v}
\end{split}
\end{equation}
and $T_{\phi\phi} = T_{\theta\theta} \sin^2\theta$.

Inserting the renormalized energy-momentum tensor into the semiclassical field equations in (\ref{semiclassial field eqs}), we find two constraint equations,
\begin{equation} \label{constraint eqs}
\begin{split}
r_{,uu} 
&= 2 \sigma_{,u} r_{,u} - 4\pi r \left[ T_{uu} 
+ \frac{P}{4\pi r^2}
(\sigma_{,uu} - \sigma_{,u}^2) \right]
\\
r_{,vv} 
&= 2 \sigma_{,v} r_{,v} - 4\pi r \left[ T_{vv} 
+ \frac{P}{4\pi r^2}
(\sigma_{,vv} - \sigma_{,v}^2) \right],
\end{split}
\end{equation}
and two evolution equations,
\begin{align} 
\sigma_{,uv} 
&= \frac{1}{4r^2} \left[4 r_{,u} r_{,v} + e^{2\sigma} 
- 8\pi \left( 2 r^2 T_{uv} +  e^{2\sigma}   T_{\theta\theta} \right)
\right]
\notag
\\
&\qquad\quad 
\times
\left( 1 - \frac{P}{r^2} \right)^{-1}
\label{sigma r evo}
\\
r_{,uv} 
&= - \frac{1}{4r} \left[4 r_{,u} r_{,v} + e^{2\sigma} 
- 16\pi r^2 \left( T_{uv} 
- \frac{P}{4\pi r^2}
\sigma_{,uv} \right) \right].
\notag
\end{align}
Note that the $\sigma$ evolution equation is divergent for $r \rightarrow \sqrt{P}$. As with the analogous divergence in the static equations, this divergence follows from the choice of the multiplicative factor in (\ref{Polyakov approx}). This divergence will play a role in our simulations and we make the not uncommon interpretation that this divergence corresponds to the central singularity, which has been shifted from $r = 0$ by semiclassical effects \cite{Sorkin:2001hf, Fabbri:2005mw, Hong:2008mw, Hwang:2010im}. We therefore require $\sqrt{P}$ to be small compared to any length scale in the system, such as the radius of a black hole horizon.

The Misner-Sharp mass function, $m(u,v)$, is defined by
\begin{equation}
g^{\mu\nu} r_{,\mu} r_{,\nu}
= 1 - \frac{2 m}{r},
\end{equation}
which leads to
\begin{equation} \label{mass function}
m = \frac{r}{2} \left( 1 +  4e^{-2\sigma} r_{,u} r_{,v} \right),
\end{equation}
which gives the total mass inside a sphere of radius $r(u,v)$.

The evolution equations and the mass function are invariant under transformations of the form
\begin{equation} \label{coord gauge transf}
\begin{split}
u &\rightarrow \tilde{u} = \tilde{u}(u)
\\
v &\rightarrow \tilde{v} = \tilde{v}(v)
\\
\sigma &\rightarrow \tilde{\sigma} = \sigma 
- \frac{1}{2} \ln \left(\partial_u \tilde{u}\right)
- \frac{1}{2} \ln \left(\partial_v \tilde{v}\right)
\end{split}
\end{equation}
and $r$ and $\phi$ unchanged. This is a coordinate gauge transformation and $\sigma$ is a coordinate gauge field. In principle, we can use this gauge transformation to set $\sigma$ to anything we would like. One of the numerical methods we use for solving the system of equations, which we outline below, will make heavy use of this gauge transformation. For this numerical method, the $uu$ component of the renormalized energy-momentum tensor in (\ref{RSET}) transforms nontrivially under the gauge transformation. As a consequence, the top constraint equation in (\ref{constraint eqs}) also transforms nontrivially. We do not use this constraint equation in solving for our dynamical solutions, but we do use it for testing our code. Additional details are given in the Appendix.


\subsection{Initial data}

The computational domain is a two-dimensional grid of $(u,v)$ values in the ranges $u_i \leq u \leq u_f$ and $v_i \leq v \leq v_f$. We choose to set $u_i = 0$ and $v_i = 0$, but will continue to write $u_i$ and $v_i$ for completeness. $u = u_i$ and $v = v_i$ are the initial hypersurfaces, for which we must supply initial data.

The initial data will be a static quantum corrected Schwarzschild black hole, as described in Sec.~\ref{sec:static}, sometimes augmented with a pulse of scalar field. The static solutions are functions of $x$ and contain a null curvature singularity at $x \rightarrow -\infty$. This singularity will not be included in the initial data, since we will not include $x\rightarrow -\infty$, which in our computational domain corresponds to $u\rightarrow \infty$. Moreover, since the singularity is null, no additional boundary condition would be necessary on the null initial hypersurface.

We are at liberty to choose the value of $x$ at the origin of the computational domain, $x_0 \equiv x (u_i, v_i)$. With this choice, the value of $x$ at any point on either initial hypersurface is given by
\begin{equation} \label{x u v}
\begin{split}
x(u,v_i) &= x_0 - \frac{1}{2} (u - u_i)
\\
x(u_i,v) &= x_0 + \frac{1}{2} (v - v_i),
\end{split}
\end{equation}
which we use to determine, in the absence of a scalar field pulse, the values of $\sigma$ and $r$ on the initial hypersurfaces.

We will always place the wormhole throat at the origin of the computational domain, so that $x_0 = x_\text{th}$. This is convenient, since then the $u = u_i$ initial hypersurface is outside the wormhole and the $v = v_i$ initial hypersurface is inside the wormhole.

For the pulse of scalar field, we have in mind classical matter that we might try to fire into the wormhole. We will therefore place the pulse on the $u = u_i$ hypersurface, which is outside the wormhole, and $\phi$ will be zero everywhere along the $v = v_i$ hypersurface. A look at the equations in Sec.~\ref{sec:equations} shows that they depend on derivatives of $\phi$ and not on $\phi$ itself. As such, we will define the pulse in terms of $\phi_{,v}$,
\begin{equation} \label{phiV pulse}
\phi_{,v} (u_i, v) = A \sin^2 \left( \pi \frac{v - v_1}{v_2-v_1} \right)
\quad
\text{for $v_1 < v < v_2$}
\end{equation}
and zero everywhere else on the initial hypersurfaces, where $A$ is a constant. This form is commonly used because both $\phi_{,v}$ and $\phi_{,vv}$ are zero at $v = v_1$ and $v = v_2$.

When including a pulse, we choose to keep $\sigma(u_i,v)$ unchanged from the static solution, which we are at liberty to do since $\sigma$ is a coordinate gauge field. $\sigma$ and $\phi$ are then determined on the initial hypersurfaces and $r$ is determined on the $v = v_i$ hypersurface, where it is unchanged from the static solution. It remains to determine $r$ on the $u = u_i$ hypersurface. We can do this by solving the bottom constraint equation in (\ref{constraint eqs}). In this equation, $\sigma_{,v}$ and $\sigma_{,vv}$ are unchanged from the static solution as are $r(u_i, v_i)$ and $r_{,v}(u_i, v_i)$. We can solve the constraint equation numerically by integrating outward from $v = v_i$ along the $u = u_i$ hypersurface.


\subsection{Numerical methods}

To dynamically evolve the system, we solve numerically the evolution equations in (\ref{phi evo}) and (\ref{sigma r evo}) using a standard second-order predictor-corrector scheme (for a description of the scheme, see \cite{Eilon:2015axa}). Our code solves for field values at all grid points on a $u = \text{constant}$ ``row," starting at $v = v_i + \Delta v$, where $\Delta v$ is the step size between grid points, and ending at $v = v_f$, before moving to the next row.

This method exhibits second-order convergence and is sufficient for determining the stability of the static solutions. As we will see, apparent horizons form and we will be interested in accurately computing an apparent horizon out to large values of $v$. Doing so requires improved numerical accuracy. We use Eilon and Ori's adaptive gauge method, which we find to be highly efficient \cite{Eilon:2015axa}.


\subsubsection{Adaptive gauge method}
\label{sec:agm}

The adaptive gauge method makes use of the coordinate gauge freedom of the system. In each row, the value of $\sigma(u,v_i)$ is chosen such that the maximal value of $\sigma$ on the row is equal to zero. Since $\sigma$ is a coordinate gauge field, this is perfectly consistent. Eilon and Ori refer to this as $\sigma$ gauge. For our system, the maximum value always occurs at the edge of the computational domain at $v = v_f$. Making this gauge choice has the effect of increasing the number of rows near the event horizon.

In $\sigma$ gauge, the $u$ coordinates for the grid points are different than the $u$ coordinates in the original gauge, where in the original gauge the values of $\sigma(u,v_i)$ are equal to those from the static solution. Our code maintains a uniform grid of $u$ coordinates in $\sigma$ gauge. The $u$ coordinates in the original gauge, for the same grid points, are nonuniform. In this way, many $u$ coordinates are used near horizons. Indeed, for a uniform step size in $\sigma$ gauge equal to $\Delta \tilde{u} = 0.01$, the step size in the original gauge can become $\Delta u \sim 10^{-10}$ or even smaller. The $v$ coordinates of the grid points will always be uniform.

The relationship between the $u$ coordinates in $\sigma$ gauge and in the original gauge follows from the bottom equation in (\ref{coord gauge transf}). We have implemented a couple of different ways for computing the $u$ coordinates in the original gauge. One way is to integrate (\ref{coord gauge transf}), giving
\begin{equation} \label{u orig u sigma}
\int e^{2\sigma(u, v_i)} du
= \int e^{2\tilde{\sigma}(\tilde{u}, v_i)} d\tilde{u},
\end{equation}
where $\tilde{u}$ and $\tilde{\sigma}$ are the values in $\sigma$ gauge. The right-hand side is computed as our code is running using the trapezoidal rule, which is second-order accurate. The left-hand side, for a range of $u$ values, is computed beforehand for the initial data. Given the value of the right-hand side, we can determine the value of $u$ that gives the left-hand side using a standard interpolation method (e.g.~a spline). Another way is to write the integral as
\begin{equation}
u = \int e^{2(\tilde{\sigma} - \sigma)} d\tilde{u}
\end{equation}
and then to write down a formal solution using the trapezoidal rule. We then search for the value of $u$ which solves the formal solution using the Newton-Raphson method. We find that both methods work well. The results presented in this paper make use of the first method.

With the $u$ coordinates in the original gauge we can compute the remaining fields on the $v = v_i$ initial hypersurface for $\sigma$ gauge. For us, this is just $r$ (since $\phi = 0$ on the $v = v_i$ hypersurface). $r$ is gauge invariant, so the value of $r$ we use in $\sigma$ gauge is the value of $r$ for the corresponding $u$ coordinate in the original gauge.


\section{Results}
\label{sec:results}

\begin{figure*}
\includegraphics[width=7in]{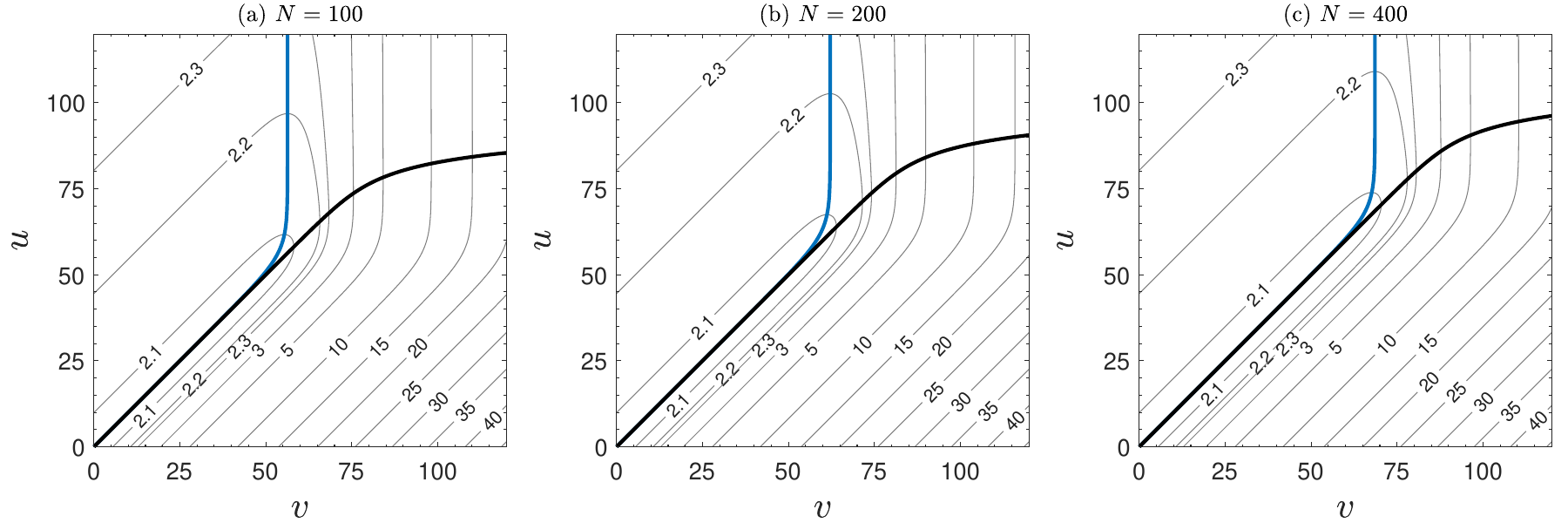} 
\caption{Dynamical evolution of a static quantum corrected Schwarzschild black hole with $M = 1$ and $P = 0.1$. The gray lines are contours for the areal radius, $r$. The thick black line is an apparent horizon defined by $r_{,u} = 0$ and the thick blue line is an apparent horizon defined by $r_{,v} = 0$. There is no scalar field pulse included and the perturbation is from discretization error alone. We find that the wormhole is expanding. Each plot uses the same initial data, but evolves the system using a uniform grid with grid spacing $\Delta u = \Delta v = 1/N$, where (a) $N = 100$, (b) 200, and (c) 400. As $N$ increases, the strength of the perturbation decreases and the static solution holds its configuration longer. This is the expected behavior for an unstable static solution.}
\label{fig:no pulse}
\end{figure*}

In Fig.~\ref{fig:no pulse}(a), we show the dynamical evolution of a static solution with $M = 1$, $P = 0.1$, and no scalar field pulse. The perturbation in Fig.~\ref{fig:no pulse}(a) is from discretization error alone. The thin gray lines are contour lines for the areal radius, $r$. The thick black and blue lines are apparent horizons, defined by $r_{,u} = 0$ and $r_{,v} = 0$, respectively.  

We recall that the wormhole throat of the static solution is located at the origin, that the region to the right of the origin is outside the wormhole and is where we reside, and the region above the origin is inside the wormhole. The region above contains relatively few contour lines. We can understand why by looking at Fig.~\ref{fig:static solutions}(a), where we see that the solid blue curve passes through a relatively small range of $r$ values as we move left from the wormhole throat.

Moving away from the lower left corner of Fig.~\ref{fig:no pulse}(a), we find apparent horizons along the wormhole throat. This is expected, since from Fig.~\ref{fig:static solutions}(a) we can see that the wormhole throat in the static solution is defined by $\partial_x r = 0$. From (\ref{u v t x}), the wormhole throat in the static solution is then also defined by $r_{,u} = r_{,v} = 0$. 

At around $u \approx 50$, the two apparent horizons separate. It is at this point that the system begins evolving away from the static solution. In between the apparent horizons, timelike and null directions necessarily find increasing values of the areal radius:~the wormhole throat is \textit{expanding}. Given that the system is evolving away from the static solution, it is not surprising that we find expansion, since the only energy-momentum in the system is from the renormalized energy-momentum tensor. For comparison, the Ellis-Bronnikov wormhole can also exhibit expansion and the analogous diagram looks similar to Fig.~\ref{fig:no pulse}(a) (cf.~figure 7(a) in \cite{Doroshkevich:2008xm}).

Since the only perturbation is from discretization error, we can decrease the size of the perturbation by decreasing the spacing between grid points. Figure \ref{fig:no pulse}(a) is made with a uniform grid with grid spacing $\Delta u = \Delta v = 1/N$ and $N = 100$. Figures \ref{fig:no pulse}(b) and \ref{fig:no pulse}(c) are evolutions with the same initial data as Fig.~\ref{fig:no pulse}(a), but with $N = 200$ and 400 respectively. As the discretization error decreases, we find that the wormhole throat takes longer before expanding. In other words, if we decrease the strength of the perturbation, the static solution is able to hold its configuration longer. This is precisely the expected behavior for an unstable static solution. Indeed, the same behavior was seen for the Ellis-Bronnikov wormhole \cite{Shinkai:2002gv}, which is known to be unstable \cite{Gonzalez:2008wd}. We conclude that this static quantum corrected Schwarzschild black hole solution is unstable.

We now introduce a scalar field pulse as an explicit perturbation. As previously mentioned, we have in mind that we are firing into the wormhole some classical matter. We make use of a simple measure for the strength of the pulse, defined as follows. We compute the mass of the system along the $u = u_i$ hypersurface, $m(u_i,v)$, using (\ref{mass function}). At large $v$, the mass approaches a constant value which, in general, is larger than the ADM mass $M$ used for the static solution (and is equal to $M$ in the absence of a pulse). We use the percent increase of the mass at large $v$, with respect to $M$, as a measure of the strength of the pulse.

\begin{figure} 
\includegraphics[width=3in]{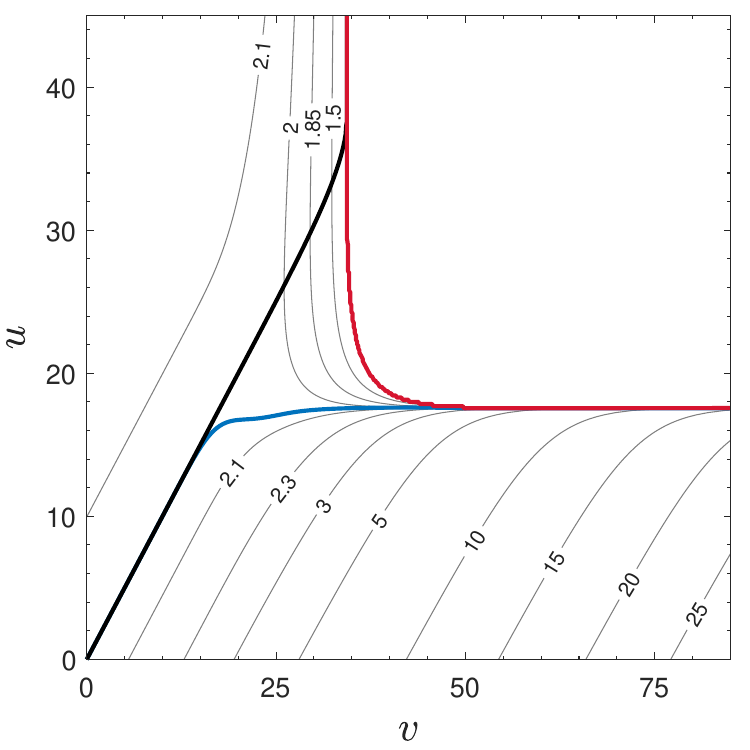} 
\caption{Dynamical evolution of a static quantum corrected Schwarzschild black hole with $M = 1$, $P = 0.1$, and a scalar field pulse as an explicit perturbation. The gray lines are contours for $r$ and the thick black and blue lines are apparent horizons, just as in Fig.~\ref{fig:no pulse}. The thick red line marks the central singularity and the region beyond the red line is not part of the spacetime. The inclusion of classical matter in the form of a scalar field pulse causes the wormhole to collapse and to form event horizons and a central singularity.}
\label{fig:pulse}
\end{figure}

In Fig.~\ref{fig:pulse}, we show results for a static solution with $M=1$, $P = 0.1$, and a pulse with parameters $A = 0.0035$, $v_1 = 10$, and $v_2 = 30$. This pulse increases the mass by roughly 1\%. As in Fig.~\ref{fig:no pulse}, the gray lines are contour lines for $r$ and the thick black and blues lines are apparent horizons. Starting in the lower left corner, we find similar behavior when compared to Fig.~\ref{fig:no pulse}, in that the apparent horizons lie along the wormhole throat and the static solution is holding its configuration. At around $u \approx 17$, the evolution moves away from the static configuration. In between the apparent horizons, timelike and null directions now find decreasing values of the areal radius:~the wormhole throat is \textit{collapsing}.

The contour lines and the apparent horizons show the formation of event horizons, one at around $u\approx 17.5$ and another at around $v \approx 34$. We therefore find that, as the wormhole collapses, black holes form over the mouths of the wormhole.  Eventually the wormhole throat collapses to $r_\text{th} \approx \sqrt{P}$, at which point we reach the central singularity, which we have indicated with the red line. The region beyond the red line is not part of the spacetime. For comparison, the Ellis-Bronnikov wormhole also exhibits collapse and the analogous diagrams look similar to Fig.~\ref{fig:pulse} (cf.~figures 2(a) and 5(a) in \cite{Doroshkevich:2008xm}).

To study the region outside the wormhole, we focus on the apparent horizon defined by $r_{,v} = 0$, which is the thick blue line in Fig.~\ref{fig:pulse}. From (\ref{mass function}), the mass inside an apparent horizon is always equal to $r/2$, where $r$ is the radius of the horizon. Since the apparent horizon aligns with the event horizon at large $v$, the mass inside the event horizon at large $v$ is equal to $r/2$.

\begin{figure} 
\includegraphics[width=3in]{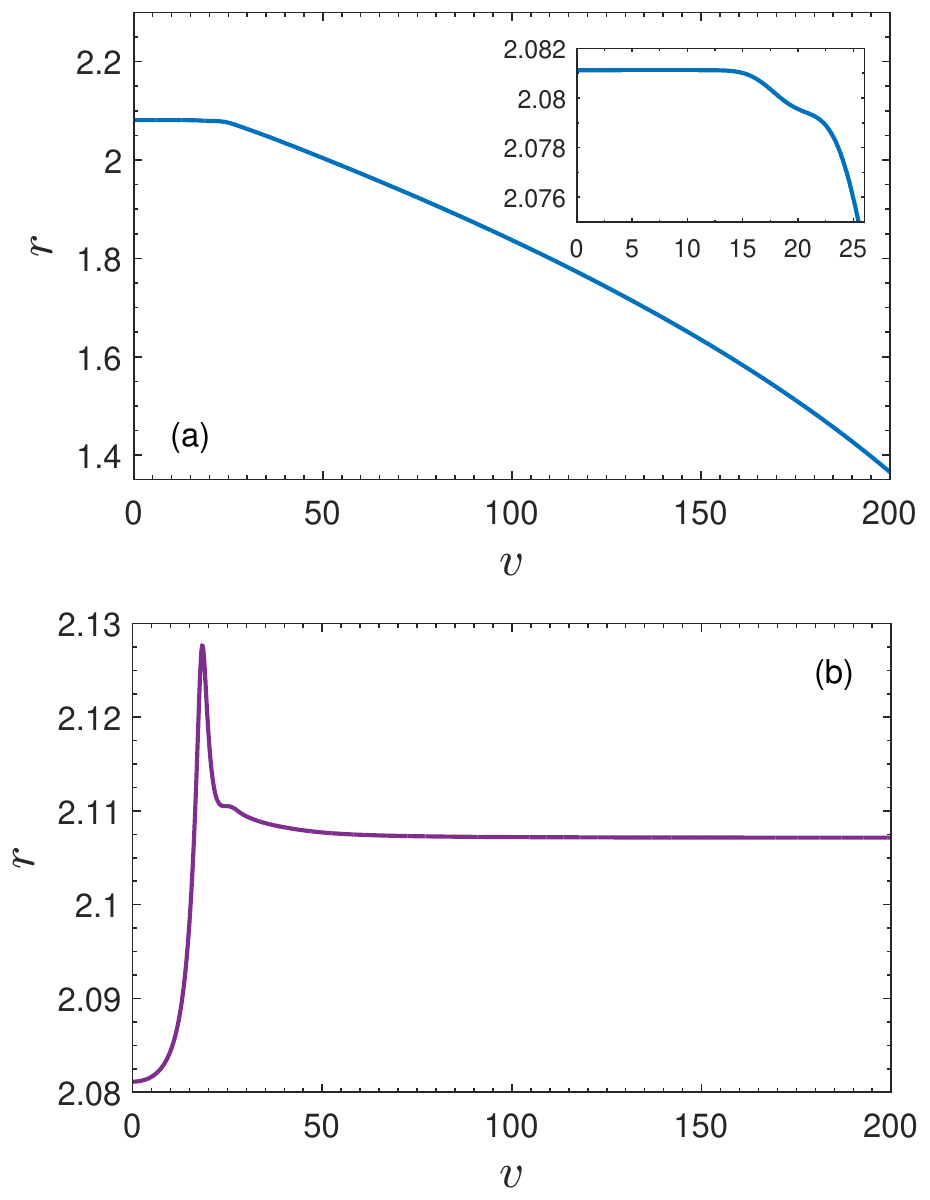} 
\caption{(a) The same apparent horizon (defined by $r_{,v} = 0$) shown as the blue curve in Fig.~\ref{fig:pulse}, but computed out to large $v$. (b) The apparent horizon computed using the same initial data as Fig.~\ref{fig:pulse}, but with evolution equations with $P = 0$. This is an inconsistent evolution and should not be taken too seriously, but it shows that with $P = 0$ the apparent horizon asymptotically approaches a constant value indicating the system evolves to a static black hole. This strongly suggests that the radius of the apparent horizon in (a) is decreasing at large $v$ because of black hole evaporation.}
\label{fig:AR}
\end{figure}

To compute the apparent horizon out to large values of $v$, we use the adaptive gauge method, as described in Sec.~\ref{sec:agm}. In Fig.~\ref{fig:AR}(a), we show the radius of the apparent horizon as a function of $v$. In the inset, we zoom into the small $v$ region. We can see that the apparent horizon starts out at $r = r_\text{th} = 2.081$. It holds its configuration until around $v \approx 15$ and then starts decreasing. It continues to decrease out to large $v$. Typically, the radius of an apparent horizon cannot decrease. There are two reasons at play here, which allow the radius to decrease. The first reason is that wormholes, which violate the null energy condition, can have apparent horizons with decreasing radii when collapsing. Such behavior occurs for the Ellis-Bronnikov wormhole \cite{Gonzalez:2008xk}. The second reason is that the renormalized energy-momentum tensor allows for black hole evaporation, during which the radius of the apparent horizon can decrease \cite{Parentani:1994ij, Ayal:1997ab}.

In the absence of the renormalized energy-momentum tensor, our expectation is that the outside of a collapsing wormhole will eventually settle down to a static Schwarzschild black hole. In this case, the radius of the apparent horizon would asymptotically approach a constant value, which would be the radius of the black hole. This is precisely what was found with the Ellis-Bronnikov wormhole \cite{Gonzalez:2008xk}. With the inclusion of the renormalized energy-momentum tensor, this cannot happen because the black hole will evaporate and therefore cannot settle down to Schwarzschild. 

To gain some insight, we consider the following \textit{inconsistent} evolution:~We use the same initial data as used in Figs.~\ref{fig:pulse} and \ref{fig:AR}(a), which are based on a static solution with $P = 0.1$ and an initial pulse computed using $P = 0.1$. We then evolve the system, but we set $P = 0$ in the evolution equations. Since the evolution equations and the initial data are inconsistent with one another, we should not take the results too seriously. Nevertheless, as shown in Fig.~\ref{fig:AR}(b), the radius of the apparent horizon asymptotically approaches a constant value at large $v$. This strongly suggests that it is black hole evaporation, as caused by the renormalized energy-momentum tensor, which accounts for the decreasing radius at large $v$ in Fig.~\ref{fig:AR}(a).

The collapse shown in Fig.~\ref{fig:pulse} occurs for a scalar field pulse that increases the mass by approximately 1\%. We continue to find that the system collapses as we lower the amplitude of the pulse down to $A = 5\times10^{-5}$, which corresponds to a $0.001$\% increase in the mass. In general, we find that a relatively small amount of classical matter is needed to trigger collapse.

As mentioned in Sec.~\ref{sec:static}, increasing the mass $M$ increases the wormhole throat radius of the static solution. As the wormhole throat radius increases, it requires more computational resources to dynamically evolve the system because the size of the computational grid must increase. We have dynamically evolved static solutions up to $M = 5$ and found that, in the absence of a scalar field pulse, they are unstable and the wormhole expands. For $M = 5$ and $P = 0.1$, using a uniform grid with $\Delta u = \Delta v = 1/100$, the apparent horizons separate and the wormhole begins expanding at roughly $u \approx 300$. Aside from this, the resulting diagram looks similar to Fig.~\ref{fig:no pulse}. We also continue to find for static solutions with larger masses that a relatively small amount of classical matter triggers collapse and that the resulting diagrams look similar to Fig.~\ref{fig:pulse}.

\begin{figure*} 
\includegraphics[width=7in]{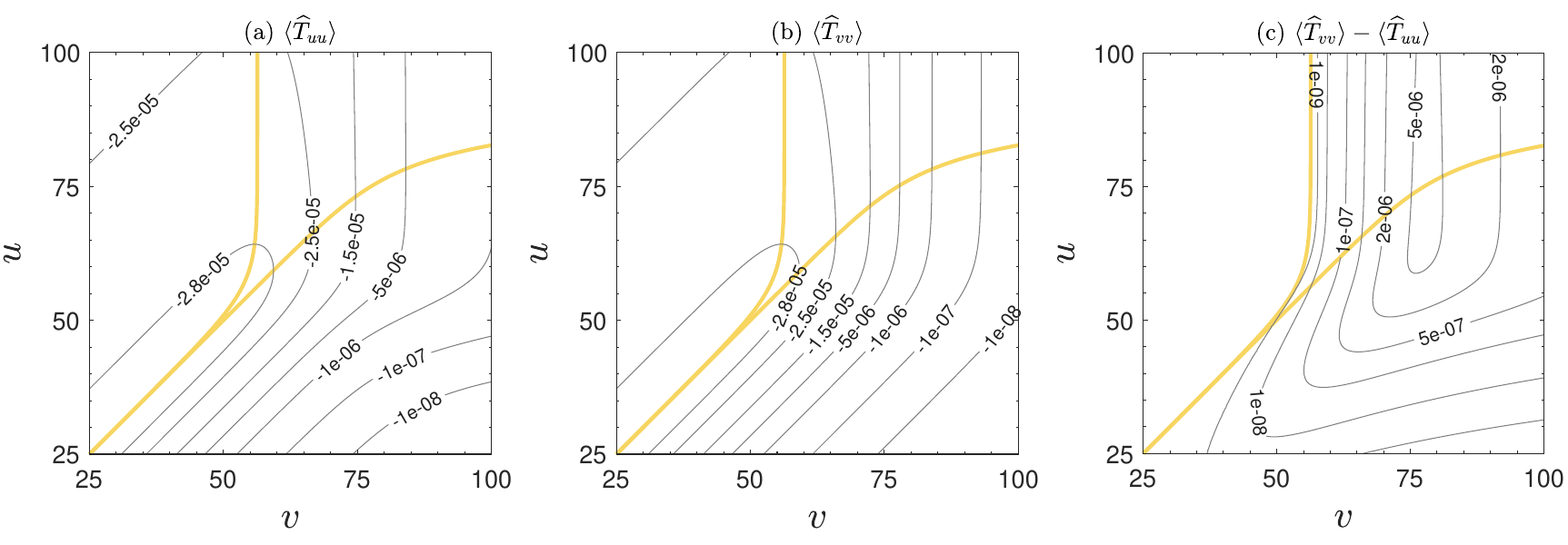} 
\caption{Contour diagrams for (a) the $uu$ component and (b) the $vv$ component of the energy-momentum tensor for the dynamical evolution shown in Fig.~\ref{fig:no pulse}(a). The yellow curves are the apparent horizons shown in Fig.~\ref{fig:no pulse}(a). The difference between the energy-momentum tensor components is displayed in (c). The region on the left side in (c) is very close to zero, with the precise values being difficult to compute numerically.}
\label{fig:TUU TVV no pulse}
\end{figure*}

\begin{figure*}
\includegraphics[width=7in]{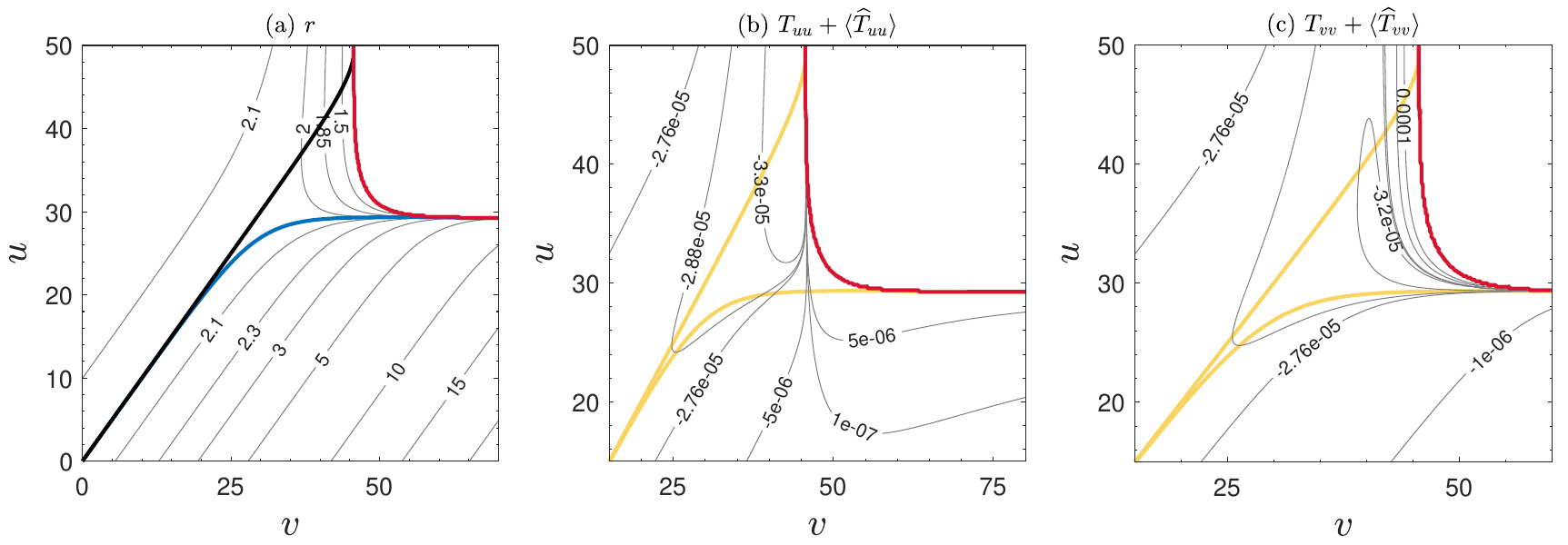} 
\caption{Dynamical evolution of a static quantum corrected Schwarzschild black hole with $M = 1$, $P = 0.1$, and a scalar field pulse with $v_1 = 5$, $v_2 = 10$, and $A = 0.0005$. (a) Contour diagram for the areal radius, $r$. The black and blue lines are apparent horizons $r_{,u} = 0$ and $r_{,v} = 0$, respectively, and the red line is the central singularity. (b) Contour diagram for the $uu$ component of the energy-momentum tensor. The apparent horizons are included for convenience as the yellow lines. (c) Contour diagram for the $vv$ component of the energy-momentum tensor. Some of the contours are unlabeled:~Starting at the upper left and moving to the right, the contour values are -2.76e-5, -2.88e-5, -3.2e-5, -2.88e-5, -2.76e-5, -1e-6, and +1e-4.}
\label{fig:TUU TVV}
\end{figure*}

We can gain some insight as to why the wormhole in Fig.~\ref{fig:no pulse} expands and why the wormhole in Fig.~\ref{fig:pulse} collapses by looking at components of the energy-momentum tensor. For the evolution shown in Fig.~\ref{fig:no pulse}(a), contour diagrams for the $uu$ and $vv$ components of the energy-momentum tensor are shown in Figs.~\ref{fig:TUU TVV no pulse}(a) and \ref{fig:TUU TVV no pulse}(b) (the apparent horizons from Fig.~\ref{fig:no pulse}(a) are shown in yellow). The $uu$ component describes outgoing energy-momentum and the $vv$ component describes ingoing energy-momentum. As can be seen in Figs.~\ref{fig:TUU TVV no pulse}(a) and \ref{fig:TUU TVV no pulse}(b), both are negative. We therefore show in Fig.~\ref{fig:TUU TVV no pulse}(c) their difference, $\langle \widehat{T}_{vv}\rangle - \langle\widehat{T}_{uu} \rangle$. Since this difference is non-negative, we have a net flow of ingoing energy-momentum. This is consistent with the mass increasing, which is easily seen to be the case along the apparent horizons in Fig.~\ref{fig:no pulse}(a), since they move along increasing radii. From the relationship between mass and the wormhole throat radius we previously reviewed for static solutions, the net flow of ingoing energy-momentum is also consistent with the wormhole expanding.

We could show analogous plots for the collapsing wormhole in Fig.~\ref{fig:pulse}. However, it is easier to see what is happening if we consider a slightly different evolution. The evolution shown in Fig.~\ref{fig:pulse} makes use of an initial pulse that is nonzero for $v_1 < v < v_2$, where $v_1 = 10$ and $v_2 = 20$. This range of $v$ values puts the initial pulse close to the interesting region where the wormhole throat is collapsing, which ends up leading to somewhat chaotic values for energy-momentum tensor in this region. Results are simplified for a pulse with $v_1 = 5$, $v_2 = 10$, and amplitude $A = 0.0005$. The contour diagram for $r$ is shown in Fig.~\ref{fig:TUU TVV}(a). We can see that it is very similar to Fig.~\ref{fig:pulse}, except that it takes longer for the wormhole to begin collapsing because the pulse is weaker. In Figs.~\ref{fig:TUU TVV}(b) and \ref{fig:TUU TVV}(c), we show the $uu$ and $vv$ components of the energy-momentum tensor (the apparent horizons from Fig.~\ref{fig:TUU TVV}(a) are shown in yellow). A horizon forms outside the wormhole at $u \approx 29$. Outside this horizon ($u\apprle 29$) and at large $v$, which approaches future null infinity, the $uu$ component in Fig.~\ref{fig:TUU TVV}(b) is positive, indicating outgoing energy-momentum, and the $vv$ component in Fig.~\ref{fig:TUU TVV}(c) is negative, also indicating outgoing energy-momentum. This net effect of outgoing energy-momentum is consistent with decreasing mass and wormhole collapse.

For a wormhole to exist, the energy-momentum tensor around the wormhole throat must violate the null energy condition \cite{Morris:1988cz}. In double null coordinates, the null energy condition is violated if $T_{uu} + \langle \widehat{T}_{uu} \rangle < 0$ or $T_{vv} + \langle \widehat{T}_{vv} \rangle < 0$. If the null energy condition is satisfied and the wormhole is collapsing, we should expect focusing of null geodesics and the formation of caustics, as follows from the Raychaudhuri equation and the focusing theorem \cite{Hawking:1973uf}. This strongly suggests the formation of a singularity. Indeed, from Figs.~\ref{fig:TUU TVV}(b) and  \ref{fig:TUU TVV}(c) we can see that the null energy condition flips from being violated to being satisfied right about where the singularity forms at $u\approx 29$ and $v \approx 45$.


\section{Conclusion}
\label{sec:conclusion}

The classical static spherically symmetric vacuum solution is the Schwarzschild black hole. When extended to include semiclassical corrections in the form of a renormalized energy-momentum tensor, the horizon disappears and is replaced by a wormhole \cite{Fabbri:2005zn, Ho:2017joh, Ho:2017vgi, Arrechea:2019jgx}. Since the renormalized energy-momentum tensor can describe black hole evaporation \cite{Parentani:1994ij, Ayal:1997ab}, it is perhaps not surprising that one does not find a \textit{static} black hole solution, since an evaporating black hole is not static.

We have studied the stability of the quantum corrected static vacuum solution by using it as the initial data of a dynamical evolution. We have shown that the static solution is unstable and that the wormhole will expand or collapse.

In the absence of classical matter, the wormhole expands since the only energy-momentum in the system is from the renormalized energy-momentum tensor. On the other hand, if there is even a small amount of classical matter present, our results indicate that the wormhole collapses, that it forms a horizon and a central singularity, and that it evolves to an evaporating black hole. Since only a small amount of classical matter is necessary to trigger collapse, it appears unlikely for the expanding wormhole to form naturally. Instead, we expect an evaporating black hole to be the astrophysically relevant system.

In our study of the collapsing dynamical solution, we purposely focused on the region outside the wormhole, which is where we reside. However, there is also the region reached by passing through the wormhole. This is properly the region on the other side of the wormhole, but for convenience we have referred to this region as inside the wormhole. This is the upper left region of Fig.~\ref{fig:pulse} and it is in this region that the static solution contains a null curvature singularity at $u\rightarrow \infty$, whose dynamical evolution deserves further study.

There may be some challenges in studying this region. We previously mentioned that only a small range of the areal radius, $r$, is probed for a relatively large range of the outgoing null coordinate, $u$. If we would like to evolve further into this region, this will require increased computational resources. However, it may be possible to compress this region using a coordinate gauge transformation. Additionally, the further we move into this region, the smaller the metric component $e^\sigma$ becomes, as can be seen from Fig.~\ref{fig:static solutions}(b), which may cause numerical challenges. In terms of the apparent horizon that is inside the wormhole (the black curve in Fig.~\ref{fig:pulse} defined by $r_{,u} = 0$), we expect that it can be studied similarly to how we studied the apparent horizon that is outside the wormhole. We would need to write code that solves the evolution equations along ``columns" instead of rows so that we could use the adaptive gauge method applied to the $v$ coordinate up until the $r_{,u} = 0$ apparent horizon is computed.

Another potential direction of study is to dynamically evolve the quantum corrected Reissner-Nordstr\"om \cite{Arrechea:2021ldl} or Einstein-Yang-Mills \cite{Kain:2025ygm} black holes. We expect nonextremal black holes of these types to also be unstable, but this remains to be confirmed. The quantum corrected Reissner-Nordstr\"om black hole has an extremal solution and it would be interesting to determine how this static solution evolves.


\acknowledgements

I am thankful to the referee, whose suggestions led to an improved paper.


\appendix

\section{Code tests}
\label{app:code tests}

\begin{figure} 
\includegraphics[width=3in]{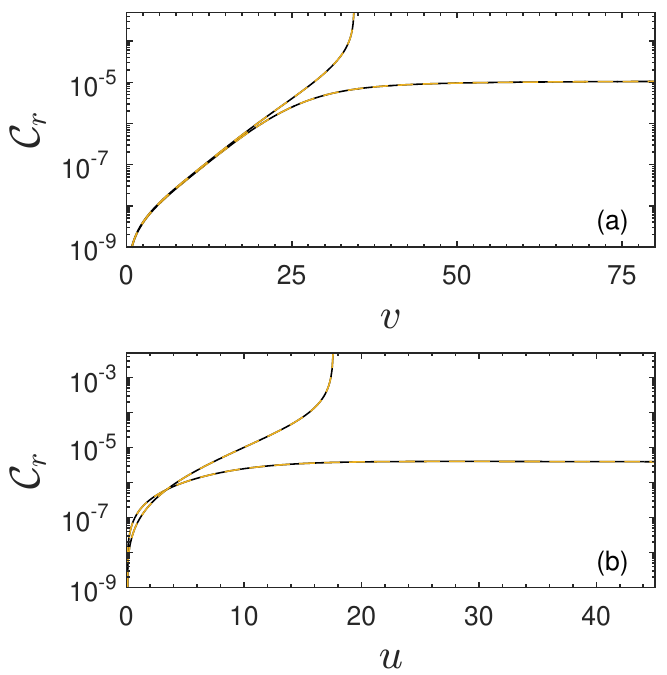} 
\caption{The convergence function in (\ref{conv fun}) is plotted for $f = r$ and with grid spacing $\Delta u = \Delta v = 1/N$ and $N = 100$, 200, and 400 where the sum is performed over (a) rows and (b) columns. These curves indicate our code is second-order convergent. See the explanation given in the Appendix for details.}
\label{fig:A1}
\end{figure}

As a first basic test of our code, we have confirmed that if we use the static classical Schwarzschild solution for initial data, our code reproduces the static solution dynamically. This is the case when using the adaptive gauge method and when not using it.

To present a test of convergence, we focus on the results shown in Fig.~\ref{fig:pulse}, which do not make use of the adaptive gauge method. Using the same initial data, we have computed the dynamical solution using three different uniform grids, defined by grid spacings $\Delta u = \Delta v = 1/N$ with $N = 100$, 200, and 400. Using these results, we compute the convergence function
\begin{equation} \label{conv fun}
\mathcal{C}_f^{N_1, N_2} = \sum_i | f_i^{N_1} - f_i^{N_2} |,
\end{equation}
where $f_i^{N}$ is the value of field $f$ computed at grid point $i$ using grid spacing $N$. In (\ref{conv fun}), $f_i^{N_1}$ and $f_i^{N_2}$ must be evaluated at the same grid points.  In Fig.~\ref{fig:A1}(a), we show results for $f = r$ where the sum is performed over rows with constant $u$. The dashed black curves are for $\mathcal{C}_r^{100, 200}$ and the dashed yellow curves are for $4C_r^{200, 400}$. The lower set of curves is for $u = 10$ and the upper set of curves is for $u = 30$. Since $C_r^{200, 400}$ is multiplied by $4$ and the curves overlap, we have second-order convergence \cite{AlcubierreBook}. This continues to be the case for the upper curves as they approach the singularity, which is why the upper curves veer upward. Figure \ref{fig:A1}(b) is an analogous plot, but summing over columns with constant $v$. The lower set of curves is for $v = 25$ and the upper set of curves is for $v = 60$. We find similar results indicating second-order convergence when using $f = \sigma$ and $f = \phi$.

Our code primarily uses the evolution equations in (\ref{phi evo}) and (\ref{sigma r evo}). We have therefore confirmed that our dynamical results satisfy both constraint equations in (\ref{constraint eqs}), as required. When not using the adaptive gauge method, we use the constraint equations as written in (\ref{constraint eqs}). When using the adaptive gauge method, the upper constraint equation in (\ref{constraint eqs}) takes on a different form. For the coordinate transformation $u \rightarrow \tilde{u} = \tilde{u}(u)$, defined by
\begin{equation}
\partial_u \tilde{u} = e^{2(\sigma - \tilde{\sigma})},
\end{equation}
the renormalized energy-momentum tensor component $\langle\widehat{T}_{uu}\rangle$ in (\ref{RSET}) undergoes a nontrivial transformation, so that the constraint equation becomes 
\begin{equation} \label{transformed con eq}
r_{,\tilde{u}\tilde{u}} 
= 2 \tilde{\sigma}_{,\tilde{u}} r_{,\tilde{u}} - 4\pi r \left[ T_{uu} 
+ \frac{P}{4\pi r^2}
(\sigma_{,\tilde{u}\tilde{u}} + \sigma_{,\tilde{u}}^2 - 2 \sigma_{,\tilde{u}} \tilde{\sigma}_{,\tilde{u}}) \right],
\end{equation}
where $T_{uu} = \phi_{,\tilde{u}}^2$ and where $\tilde{u}$ and $\tilde{\sigma}$ are the values in $\sigma$ gauge. 

\begin{figure}
\includegraphics[width=3in]{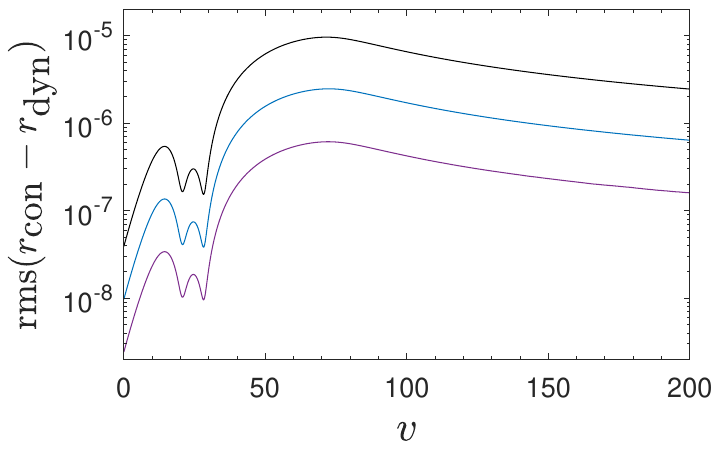} 
\caption{Equation (\ref{rms}), where $r_\text{con}$ is computed from the constraint equation in (\ref{transformed con eq}) and $r_\text{dyn}$ is from the dynamical evolution. From top to bottom, the curves are computed using grid spacing $\Delta \tilde{u} = \Delta v = 1/N$ and $N = 50$, 100, and 200. That the curves drop by a factor of $4$ when the grid spacing drops by a factor of $2$ indicates second-order convergence. That the results are small indicates that the constraint equation in (\ref{transformed con eq}) is satisfied by the dynamical evolution.}
\label{fig:A2}
\end{figure}

For an example of our code satisfying the constraint equations, we focus on the results in Fig.~\ref{fig:AR}(a), which are computed using the adaptive gauge method. In computing these results, we end the computation just after the apparent horizon is computed. As such, the computation does not reach the singularity. We solve the constraint equation in (\ref{transformed con eq}) for $r$ using second-order Runge-Kutta and second-order finite differencing for the derivatives. Note that all derivatives are in terms of $\tilde{u}$, which have uniform spacing, making the finite differencing straightforward. At each grid point, we compute the difference between the result obtained from the constraint equation and the result from the dynamical evolution. We then compute the root-mean-square (rms) value,
\begin{equation} \label{rms}
\text{rms}(r_\text{con} - r_\text{dyn}),
\end{equation}
along columns of constant $v$. The result is shown in Fig.~\ref{fig:A2} for grid spacing $\Delta \tilde{u} = \Delta v = 1/N$ and $N = 50$, $100$, and $200$. That the curves drop by a factor of $4$ when the grid spacing drops by a factor of $2$ indicates second-order convergence. That the results are small indicates that the constraint equation is satisfied.




%

\end{document}